\documentclass[aps,pra,floatfix,superscriptaddress,twocolumn]{revtex4}
\usepackage[dvips]{graphicx}

\usepackage{longtable}
\usepackage{dcolumn}
\usepackage[dvips]{graphicx}
\usepackage{bm}
\usepackage{bbm}

\usepackage{times}
\usepackage{nicefrac}
\usepackage{amsmath}
\usepackage{amsfonts}
\usepackage{amssymb}
\usepackage{amsthm}

\newcolumntype{.}{D{x}{}{-1}}

\newcommand{\vare}{\varepsilon}

\newcommand{\lbr}{\langle}
\newcommand{\rbr}{\rangle}

\newcommand{\Za}{Z\alpha}

\begin{document}

\title{Weighted difference of $\bm g$-factors of light Li-like and H-like ions for an improved determination\\
of the fine-structure constant}

\author{V.~A. Yerokhin}
\address{Max~Planck~Institute for Nuclear Physics, Saupfercheckweg~1, 69117 Heidelberg,
Germany} \address{Center for Advanced Studies, Peter the Great St.~Petersburg Polytechnic
University, 195251 St.~Petersburg, Russia}

\author{E.~Berseneva}
\address{Max~Planck~Institute for Nuclear Physics, Saupfercheckweg~1, 69117 Heidelberg, Germany}
\address{Department of Physics, St. Petersburg State University,
7/9 Universitetskaya naberezhnaya, St. Petersburg 199034, Russia}

\author{Z. Harman}
\address{Max~Planck~Institute for Nuclear Physics, Saupfercheckweg~1, 69117 Heidelberg, Germany}

\author{I.~I.~Tupitsyn}
\address{Department of Physics, St. Petersburg State University,
7/9 Universitetskaya naberezhnaya, St. Petersburg 199034, Russia}

\author{C.~H. Keitel}
\address{Max~Planck~Institute for Nuclear Physics, Saupfercheckweg~1, 69117 Heidelberg, Germany}

\begin{abstract}

A weighted difference of the $g$-factors of the Li- and H-like ion of the same element is studied
and optimized in order to maximize the cancellation of nuclear effects. To this end, a detailed
theoretical investigation is performed for the finite nuclear size correction to the one-electron
$g$-factor, the one- and two-photon exchange effects, and the QED effects. The coefficients of
the $\Za$ expansion of these corrections are determined, which allows us to set up the optimal
definition of the weighted difference. It is demonstrated that, for moderately light elements,
such weighted difference is nearly free from uncertainties associated with nuclear effects and
can be utilized to extract the fine-structure constant from bound-electron $g$-factor experiments
with an accuracy competitive with or better than its current literature value.

\end{abstract}


\maketitle

\section{Introduction}

Modern measurements of the bound-electron $g$-factor in H-like ions have reached the level of
fractional accuracy of $3\times 10^{-11}$ \cite{sturm:14}. Experiments have also been performed
with Li-like ions \cite{wagner:13}. In future it shall be possible to conduct similar experiments
not only with a single ion in the trap, but also with several ions simultaneously. Such a setup
would allow one to directly access differences of the $g$-factors of different ions, thus largely
reducing systematic uncertainties and possibly gaining about two orders of magnitude in
experimental accuracy~\cite{sturm:priv}. So, experimental investigations of differences of the
bound-electron $g$ factors on a sub-$10^{-12}$ level look feasible in the future. Such measurements
would become sensitive to the uncertainty of the fine-structure constant $\alpha$, which is
presently known up to the fractional accuracy of $3\times 10^{-10}$ \cite{mohr:12:codata}. It might
be tempting to use such future experiments as a tool for an independent determination of $\alpha$.

In order to accomplish a competitive determination of $\alpha$ from the bound-electron $g$-factor
experiments, one has to complete theoretical calculations to a matching accuracy, which is a
challenging task. One of the important problems on the way is the uncertainty due to nuclear
effects, which cannot be well understood at present. These uncertainties set a limitation on the
ultimate accuracy of the theoretical description and, therefore, on the determination of $\alpha$.

There is a way to reduce the nuclear effects and the associated uncertainties, by forming
differences of different charge states of the same element. In Ref.~\cite{shabaev:02:li}, it was
suggested to use a weighted difference of the $g$-factors of the H- and Li-like ions of the same
element in order to suppress the nuclear size effects by about two orders of magnitude for high-$Z$
ions. In Ref.~\cite{shabaev:06:prl}, a weighted difference of the $g$-factors of B-like and H-like
charge states of the same element was proposed. It was shown that the theoretical uncertainty of
the nuclear size effect for ions around Pb can be reduced to $4\times10^{-10}$, which was several
times smaller than the uncertainty due to the fine-structure constant at the time of publication of
Ref.~\cite{shabaev:06:prl}. Since then, however, the uncertainty of $\alpha$ was decreased by an
order of magnitude \cite{bouchendira:11,aoyama:12,aoyama:15}, thus making it more difficult to
access it in the bound-electron $g$-factor experiments. In our recent Letter
\cite{yerokhin:16:gfact:prl} we proposed a weighted difference of the $g$-factors of low-$Z$
Li-like and H-like ions, for which a more significant cancellation of nuclear effects can be
achieved. In the present paper we describe details of the underlying calculations and report
extended numerical results for the finite nuclear size corrections.

In our approach, the weight $\Xi$ of the specific difference of the $g$-factors is determined on
the basis of studying the $\Za$ and $1/Z$ expansions of various finite nuclear size (fns)
corrections, in such a way that the cancellation of these undesirable contributions is maximized.
We introduce the following $\Xi$-weighted difference of the bound-electron $g$-factors of the
Li-like and H-like charge states of the same element,
\begin{align} \label{eq:36}
\delta_{\Xi}g = g(2s) - \Xi\, g(1s)\,,
\end{align}
where $g(2s)$ is the $g$-factor of the Li-like ion, $g(1s)$ is the $g$-factor of the H-like ion, and the
parameter $\Xi$ is defined as
\begin{align} \label{eq:37}
\Xi = 2^{-2\gamma-1}\,\left[ 1 + \frac{3}{16}(\Za)^2\right]  \left(1-\frac{2851}{1000}\frac1Z
 + \frac{107}{100}\frac1{Z^2}\right)\,,
\end{align}
with the notation $\gamma  = \sqrt{1-(\Za)^2}$. The justification of this choice of $\Xi$ will be
given later, after studying the contributions of individual physical terms to the fns effect.

This article is organized as follows. In Section~\ref{sec:fnscorr} we describe our calculations of
various fns contributions, namely, the leading one-electron fns effect, the fns correction from the
one-electron QED effects, and the two- and three-electron fns corrections due to the exchange of
one or more photons between the electrons. The resulting weighted difference of the $g$-factors and
its utility in determining the fine-structure constant are discussed in Section~\ref{sec:wdiff},
which is followed by a short conclusion.

\section{\label{sec:fnscorr} Finite nuclear size corrections}

\subsection{One-electron finite nuclear size}
\label{sec:sub1}
The leading one-electron fns correction to the bound-electron $g$-factor is
defined as follows:
\begin{equation} \label{eq:01}
\delta g_{\rm N}^{(0)} = g_{\rm ext}^{(0)} - g_{\rm pnt}^{(0)}\,,
\end{equation}
where $g_{\rm ext}^{(0)}$ and $g_{\rm pnt}^{(0)}$ are the leading-order bound-electron $g$ factor
values calculated assuming the extended and the point-like nuclear models, respectively. The
leading-order bound-electron $g$ factor is obtained for $ns$ states as
\begin{equation} \label{eq:03}
g^{(0)} = -\frac83\, \int_0^{\infty}dr\,r^3\,g_a(r)\,f_a(r)\,,
\end{equation}
where $g_a$ and $f_a$ are the upper and the lower radial components of the $ns$ Dirac wave
function, respectively \cite{rose:61}.

The fns correction $\delta g_{\rm N}^{(0)}$ has an approximate relation to the corresponding
correction to the Dirac energy, which reads \cite{karshenboim:05} for $ns$ states as
\begin{equation} \label{eq:03a}
\delta g_{\rm N}^{(0)} = \frac43\,(2\gamma+1)\,\frac{\delta E_{\rm N}}{m}\,,
\end{equation}
where $\delta E_{\rm N}$ is the nuclear-size correction to the Dirac energy. Eq. (\ref{eq:03a}) is exact
in the nonrelativistic limit and also holds with a reasonable accuracy in the whole region of nuclear
charge numbers $Z$. Using Eq.~(\ref{eq:03a}) and the result of Ref.~\cite{shabaev:93:fns} for
$\delta E_{\rm N}$, the leading one-electron fns effect for $ns$ states can be parameterized as
\begin{align} \label{eq:04}
\delta g_{\rm N}^{(0)} = &\ \frac{2}{5}\,\left( \frac{2\,\Za\, R_{\rm sph}}{n}\right)^{2\gamma}\,
  \frac{(\Za)^2}{n}\,
%\nonumber \\ & \times
  \left[ 1 + (\Za)^2\,H_n^{(0,2+)}\right]\,,
\end{align}
where $R_{\rm sph} = \sqrt{5/3}\,R$ is the radius of the nuclear sphere with the root-mean-square
(rms) charge radius $R$ and $H_n^{(0,2+)}$ is the remainder due to relativistic effects. The
superscript $(0,2+)$ indicates that its contribution is of zeroth order in $1/Z$ and of second and
higher orders in $\Za$. The nonrelativistic limit of Eq.~(\ref{eq:04}) agrees with the well-known
result of Refs. \cite{karshenboim:00:pla,glazov:01:pla}.

The leading relativistic correction $H_n^{(0,2)}$ has been given in a closed analytical form in
Ref.~\cite{glazov:01:pla}. We deduce from it that the difference of the relativistic corrections of
relative order $(\Za)^2$ for $2s$ and $1s$ states does not depend on the nuclear charge radius nor
on the nuclear charge distribution model, and is just a constant:
\begin{align} \label{eq:05}
H_{21}^{(0,2)} \equiv H_2^{(0,2)} - H_1^{(0,2)} = \frac{3}{16} \,.
\end{align}

In the present work we calculate the nuclear-size correction $\delta g_{\rm N}^{(0)}$ numerically.
For the extended nucleus, the radial Dirac equation is solved with the Dual Kinetic Balance (DKB)
method \cite{shabaev:04:DKB}, which allows us to determine $g_{\rm ext}^{(0)}$ with a very high
accuracy. The nuclear-size correction is obtained by subtracting the analytical point-nucleus
result. In order to avoid loss of numerical accuracy in the low-$Z$ region, we used the DKB method
implemented in the quadruple (about 32 digits) arithmetics.

In our calculations, we used three models of the nuclear charge distribution. The two-parameter
Fermi model is given by
\begin{align}\label{fermi}
\rho_{\rm Fer}(r) = \frac{N}{1+\exp[(r-r_0)/a]}\,,
\end{align}
where $r_0$ and $a$ are the parameters of the Fermi distribution, and $N$ is the normalization
factor. The parameter $a$ was fixed by the standard choice of $a = 2.3/(4\ln3)\approx 0.52$~fm. The
homogeneously charged sphere distribution of the nuclear charge is given by
\begin{align} \label{1}
\rho_{\rm Sph}(r) = \frac{3}{4\pi R^3_{\rm sph}}\,\theta(R_{\rm Sph}-r)\,,
\end{align}
where $\theta$ is the Heaviside step function. The Gauss distribution of the nuclear charge reads
\begin{align} \label{1a}
\rho_{\rm Gauss}(r) = \left( \frac{3}{2\pi R^2 }\right)^{3/2}\,\exp\left(-\frac{3\, r^2}{2R^2}\right)\,.
\end{align}

The results of our calculations for the $2s$ and $1s$ states are presented in Table~\ref{tab:Z0},
expressed in terms of the function $H_n^{(0,2+)}$. Experimental values of the rms nuclear charge
radii $R$ are taken from Ref.~\cite{angeli:13}. For ions with $Z \ge 10$, we perform calculations
with the Fermi and the homogeneously charged sphere models. The difference of the values obtained
with these two models is taken as an estimation of the model dependence of the results. For ions
with $Z < 10$, the Fermi model is no longer adequate and we use the Gauss model instead.

We observe that the model dependence of the relativistic fns correction $H_n^{(0,2+)}$ is generally
not negligible; it varies from 1\% in the medium-$Z$ region to 5\% in the low-$Z$ region. However,
the model dependence of the difference $H_2^{(0,2+)}-H_1^{(0,2+)}$ is tiny. According to
Eq.~(\ref{eq:05}), it is suppressed by a small factor of $(\Za)^2$. Our calculations show that in
addition it is suppressed by a small numerical coefficient.

We conclude that both the model dependence and the $R$ uncertainty of the one-electron fns
correction can be cancelled  up to a very high accuracy by forming a suitably chosen difference.
The following weighted difference of the $2s$ and $1s$ one-electron $g$-factors cancels the
one-electron fns contributions of relative orders $(\Za)^0$ and $(\Za)^2$,
\begin{align} \label{eq:06}
\delta_{\Xi_0}g = g^{(0)}(2s) - \Xi_0\, g^{(0)}(1s)\,,
\end{align}
with the weight
\begin{align} \label{eq:07}
\Xi_0 = 2^{-2\gamma-1}\,\left[ 1 + \frac{3}{16}(\Za)^2\right]\,.
\end{align}
The one-electron fns effects in the difference $\delta_{\Xi_0}g$ arise only in the relative order
$(\Za)^4$, with a numerically small coefficient.

%%%%%%%%%%%%%%%%%%%%%%%%%%%%%%%%%%%%%%%%%%%%%%%%%%%%%%%%%%%%%%%%%%%%%%%
\begin{table*}
\caption{The relativistic fns correction, in terms of function $H_n^{(0,2+)}$ defined by Eq.~(\ref{eq:04}), for
the $2s$ state ($n = 2$) and the $1s$ state ($n = 1$), for different models of the nuclear charge distribution.
The rms charge radii $R$ and their errors are taken from the compilation of Ref.~\cite{angeli:13}.
 \label{tab:Z0} }
\begin{center}
\begin{ruledtabular}
\begin{tabular}{lll...}
                $Z$ & $R$~[fm] & Model
                & \multicolumn{1}{c}{$H_2^{(0,2+)}$}
                & \multicolumn{1}{c}{$H_1^{(0,2+)}$}
                & \multicolumn{1}{c}{$H_2^{(0,2+)}-H_1^{(0,2+)}-3/16$}
 \\
\hline\\[-9pt]
% H_n^(2+)
%                          &    2s     &    1s     &  2s-ksi*1s \\
%
6   & 2.4702(22)  & Gauss  & 0.x9296(3) & 0.x7421(3) & 0.x00003 \\
    &             & Sphere & 0.x9827(3) & 0.x7951(3) & 0.x00007 \\[2pt]
8   & 2.6991(52)  & Gauss  & 0.x9912(6) & 0.x8035(5) & 0.x0001 \\
    &             & Sphere & 1.x0408(5) & 0.x8531(5) & 0.x0002 \\[2pt]
10  & 3.0055(21)  & Fermi  & 1.x0248    & 0.x8370    & 0.x0003 \\
    &             & Sphere & 1.x0700(2) & 0.x8822(2) & 0.x0003 \\[2pt]
12  &  3.0570(16) & Fermi  & 1.x0690    & 0.x8810    & 0.x0005 \\
    &             & Sphere & 1.x1067(1) & 0.x9186(1) & 0.x0006 \\[2pt]
14  &  3.1224(24) & Fermi  & 1.x1001(1) & 0.x9118(1) & 0.x0008 \\
    &             & Sphere & 1.x1327(1) & 0.x9443(1) & 0.x0009 \\[2pt]
20  & 3.4776(19)  & Fermi  & 1.x1542(1) & 0.x9647(1) & 0.x0020 \\
    &             & Sphere & 1.x1764(1) & 0.x9868(1) & 0.x0021 \\[2pt]
25  & 3.7057(22)  & Fermi  & 1.x1843    & 0.x9934    & 0.x0034 \\
    &             & Sphere & 1.x2030(1) & 1.x0119(1) & 0.x0035 \\[2pt]
30  & 3.9283(15)  & Fermi  & 1.x2085    & 1.x0159    & 0.x0051 \\
    &             & Sphere & 1.x2246    & 1.x0319(1) & 0.x0053 \\[2pt]
35  & 4.1629(21)  & Fermi  & 1.x2297(1) & 1.x0350    & 0.x0071 \\
    &             & Sphere & 1.x2438    & 1.x0490(1) & 0.x0073 \\[2pt]
40  & 4.2694(10)  & Fermi  & 1.x2518    & 1.x0548    & 0.x0095 \\
    &             & Sphere & 1.x2652(1) & 1.x0679    & 0.x0098 \\[2pt]
45  & 4.4945(23)  & Fermi  & 1.x2714(1) & 1.x0718    & 0.x0121 \\
    &             & Sphere & 1.x2834(1) & 1.x0836(1) & 0.x0123 \\[2pt]
50  & 4.6519(21)  & Fermi  & 1.x2920    & 1.x0897    & 0.x0148 \\
    &             & Sphere & 1.x3033(1) & 1.x1006    & 0.x0151 \\[2pt]
55  & 4.8041(46)  & Fermi  & 1.x3129(1) & 1.x1077    & 0.x0177 \\
    &             & Sphere & 1.x3235(1) & 1.x1180(1) & 0.x0180 \\[2pt]
60  & 4.9123(25)  & Fermi  & 1.x3346(1) & 1.x1265    & 0.x0206 \\
    &             & Sphere & 1.x3447    & 1.x1363(1) & 0.x0209 \\[2pt]
\end{tabular}
\end{ruledtabular}
\end{center}
\end{table*}

\subsection{One-electron QED fns correction}

The one-electron QED fns correction $\delta g_{\rm NQED}^{(0)}$ to the bound-electron $g$ factor can be
conveniently parameterized by means of the dimensionless function $G^{(0)}_{\rm NQED}$ \cite{yerokhin:13:jpb},
\begin{equation} \label{eq:qed1}
\delta g_{\rm NQED}^{(0)} = \delta g_{\rm N}^{(0)}\, \frac{\alpha}{\pi}\, G^{(0)}_{\rm NQED}(\Za,R)\,,
\end{equation}
where $\delta g_{\rm N}^{(0)}$ is the leading-order fns correction discussed in
Sec.~\ref{sec:sub1}, and $G^{(0)}_{\rm NQED}$ is a slowly varying function. The correction can be
divided into four parts,
\begin{equation}
G^{(0)}_{\rm NQED} = G_{\rm NSE} + G_{\rm NUe,el} + G_{\rm NWK,el} + G_{\rm NVP, ml}\,,
\end{equation}
where $G_{\rm NSE}$ is the contribution of the electron self-energy, $G_{\rm NUe,el}$ is induced by
the insertion of the Uehling potential into the electron line, $G_{\rm NWK,el}$ is the analogous
correction by the Wichmann-Kroll potential, and $G_{\rm NVP, ml}$ is the so-called magnetic-loop
vacuum-polarization correction.

The QED fns correction was studied in detail in our previous investigation \cite{yerokhin:13:jpb},
where we reported numerical results for the $1s$ state of H-like ions. In the present work, we
extend our calculations to the $2s$ state, which is required for describing the Li-like ions. The
numerical results obtained for the $2s$ state are listed in Table~\ref{tab:qed_fns}. The results
for the $1s$ state are taken from Ref.~\cite{yerokhin:13:jpb}. We observe that the QED fns
corrections for the $1s$ and $2s$ states, expressed in terms of the function $G^{(0)}_{\rm NQED}$,
are very close to each other. Therefore, they largely cancel in the weighted difference
$\delta_{\Xi_0}g$ introduced in Eq.~(\ref{eq:06}).

%%%%%%%%%%%%%%%%%%%%%%%%%%%%%%%%%%%%%%%%%%%%%%%%%%%%%%%%%%%%%%%%%%%%%%%
\begin{table*}
\caption{One-electron QED fns corrections to the bound-electron $g$ factor, expressed in terms of
$G^{(0)}_{\rm NQED}$ defined by Eq.~(\ref{eq:qed1}). The abbreviations are as follows: "NSE" denotes the
self-energy contribution, "NUe,el" denotes the Uehling electric-loop vacuum-polarization correction, "NWK,el"
stands for the Wichmann-Kroll electric-loop vacuum-polarization correction, and "NVP,ml" denotes the
magnetic-loop vacuum-polarization contribution.
 \label{tab:qed_fns} }
\begin{center}
\begin{ruledtabular}
\begin{tabular}{l......}
                $Z$
                & \multicolumn{1}{c}{NSE}
                & \multicolumn{1}{c}{NUe,el}
                & \multicolumn{1}{c}{NWK,el}
                & \multicolumn{1}{c}{NVP,ml}
                                         & \multicolumn{1}{c}{Total, $2s$}
                                         & \multicolumn{1}{c}{Total, $1s$} \\
\hline\\[-5pt]
  6 & -0.x54\,(20)  &   0.x179       & -0.x011        & -0.x010\,(1)    & -0.x38\,(20) & -0.x60\,(1)   \\
  8 & -0.x77\,(10)   &   0.x256       & -0.x019        & -0.x010\,(1)   & -0.x55\,(10) & -0.x70\,(1)   \\
 10 & -0.x94\,(4)   &   0.x337       & -0.x028        & -0.x013\,(1)    & -0.x65\,(4)  & -0.x807\,(9)  \\
 12 & -1.x14\,(4)   &   0.x430       & -0.x040        & -0.x017\,(2)    & -0.x77\,(4)  & -0.x905\,(8)      \\
 14 & -1.x32\,(4)   &   0.x530       & -0.x053        & -0.x018\,(2)    & -0.x86\,(4)  & -0.x996\,(5)     \\
 20 & -1.x86\,(4)   &   0.x863       & -0.x098        & -0.x025\,(4)    & -1.x12\,(4)  & -1.x237\,(3)    \\
 25 & -2.x36\,(4)   &   1.x185       & -0.x143        & -0.x030\,(4)    & -1.x35\,(4)  & -1.x404\,(2)    \\
 30 & -2.x82\,(4)   &   1.x543       & -0.x191        & -0.x035\,(6)    & -1.x50\,(4)  & -1.x542\,(2)     \\
 35 & -3.x27\,(2)   &   1.x933       & -0.x240        & -0.x039\,(8)    & -1.x62\,(4)  & -1.x655\,(1)  \\
 40 & -3.x75\,(2)   &   2.x376       & -0.x295        & -0.x044\,(8)    & -1.x71\,(2)  & -1.x733\,(1)     \\
 45 & -4.x23\,(1)   &   2.x837       & -0.x345        & -0.x047\,(10)   & -1.x79\,(2)  & -1.x793\,(1)   \\
 50 & -4.x73\,(1)   &   3.x348       & -0.x398        & -0.x050\,(12)   & -1.x83\,(1)  & -1.x821\,(1)   \\
 55 & -5.x25\,(1)   &   3.x902       & -0.x450        & -0.x053\,(12)   & -1.x85\,(1)  & -1.x819\,(1)  \\
 60 & -5.x79\,(2)   &   4.x515       & -0.x502\,(1)   & -0.x055\,(14)   & -1.x83\,(2)  & -1.x780\,(1)   \\
\end{tabular}
\end{ruledtabular}
\end{center}
\end{table*}

\subsection{One-photon exchange fns correction}

The one-photon exchange fns correction is the dominant two-electron contribution to the total fns
effect. It is suppressed by the factor of $1/Z$ with respect to the leading one-electron fns
contribution $\delta g_{\rm N}^{(0)}$. The one-photon exchange fns correction can be obtained as a
difference of the one-photon exchange contributions to the $g$-factor evaluated with the extended
nuclear charge distribution and with the point nucleus,
\begin{equation} \label{eq:21}
\delta g_{\rm N}^{(1)} = \delta g_{\rm ext}^{(1)} - \delta g_{\rm pnt}^{(1)}\,.
\end{equation}

The one-photon exchange correction to the $g$-factor of the ground and valence-excited states of
Li-like ions is given by~\cite{shabaev:02:li}
\begin{align} \label{eq:22}
    \delta g^{(1)}  = &\ 2 \sum_{\mu_c}\sum_P (-1)^P \Bigl[
    \lbr Pv\,Pc| I (\Delta_{Pc\,c})| \delta^{(1)}_Vv \,c\rbr
  \nonumber \\ & {}
 +  \lbr Pv\,Pc| I (\Delta_{Pc\,c})| v\,\delta^{(1)}_Vc \rbr
 \Bigr]
  \nonumber \\ & {}
   - \sum_{\mu_c} \bigl[\lbr v|V_g|v\rbr-\lbr c|V_g|c\rbr\bigr]\,  \lbr cv| I^{\prime}(\Delta_{vc})|vc\rbr
\,,
\end{align}
where $v$ and $c$ denote the valence and the core electron states, respectively, $\mu_c$ is the
momentum projection of the core electron, $P$ is the permutation operator, $(PvPc) = (vc)$ or
$(cv)$, $(-1)^P$ is the sign of the permutation, $\Delta_{ab} = \vare_a-\vare_b$, $I(\omega)$ is
the relativistic operator of the electron-electron interaction defined below, and $I'(\omega_0) =
dI(\omega)/(d\omega)$ at $\omega = \omega_0$. Further notations used in Eq.~(\ref{eq:22}) are as
follows: $\delta^{(1)}_Va$ stands for the first-order perturbation of the wave function $a$ by the
potential $V_g$,
\begin{align} \label{eq:23}
    |\delta^{(1)}_V a\rbr = \sum_{n}^{\vare_a\neq\vare_n}
    \frac{|n\rbr\,\lbr n|V_g|a\rbr}{\vare_a-\vare_n}\,,
\end{align}
and $V_g$ is the effective $g$-factor potential (see, e.g., Eq.~(14) of
Ref.~\cite{yerokhin:10:sehfs}),
\begin{align} \label{eq:24}
V_g(\bm{r}) = 2\,m\,[\bm{r}\times\bm{\alpha}]_z\,,
\end{align}
where $\bm{\alpha}$ is the vector of Dirac matrices in the standard representation. The above form of the
potential $V_g(\bm{r})$ assumes that the momentum projection of the valence state $v$ in Eq.~(\ref{eq:22})
is fixed as $\mu_v = 1/2$.

The relativistic electron-electron interaction operator $I(\omega)$ in the Feynman gauge reads
\begin{align} \label{eq:25}
  I(\omega,\bm{r}_1,\bm{r}_2) = \alpha\,
   \left( 1-\bm{\alpha}_1 \cdot \bm{\alpha}_2 \right)\,
                \frac{\exp\left[i|\omega|r_{12}\right]}{r_{12}} \,,
\end{align}
where $r_{12} = |\bm{r}_1-\bm{r}_2|$ is the distance between the two electrons and $\omega$ is the
frequency of the photon exchanged between them.

The calculation of the one-photon exchange contribution with the extended and the point nuclear
models was reported in Ref.~\cite{shabaev:02:li}. In the present work, we redo these calculations
with an enhanced precision, which is necessary for an accurate identification of the fns effect.
The one-photon exchange fns correction $\delta g^{(1)}_N$ can be parameterized as
\begin{align} \label{eq:26}
 \delta g^{(1)}_N = \delta g^{(0)}_N\, \frac1Z \,H^{(1)}(\Za,R)\,,
\end{align}
where $\delta g^{(0)}_N$ is the one-electron nuclear-size correction introduced earlier, and $H^{(1)}$
is a slowly varying function. The $\Za$ expansion of $H^{(1)}$ reads
\begin{align} \label{eq:26a}
H^{(1)} =  H^{(1,0)} + (\Za)^2 \,H^{(1,2+)}\,,
\end{align}
where $H^{(1,0)}$ is the leading nonrelativistic contribution and $H^{(1,2+)}$ is the higher-order
remainder.

The nuclear-size correction is evaluated in this work as the difference of Eq.~(\ref{eq:22})
calculated with the extended vs. point-like nuclear models. The numerical evaluation of
Eq.~(\ref{eq:22}) with the extended nucleus is performed by using the DKB method
\cite{shabaev:04:DKB}. For the point nucleus, we use the analytical expressions for the
reference-state wave functions and for the diagonal (in $\kappa$) $g$-factor perturbed wave
function \cite{shabaev:03:PSAS}, and the standard implementation of the $B$-splines method
\cite{johnson:88} for the non-diagonal in $\kappa$ part of the perturbed wave function. In order to
avoid loss of numerical accuracy in the low-$Z$ region, we employ the DKB and the $B$-splines
methods implemented in the quadruple arithmetics.

The accuracy of the obtained numerical results is checked as follows. We observe that the leading
term of the $\Za$ expansion of Eq.~(\ref{eq:26a}), $H^{(1,0)}$, should not depend on the nuclear
charge radius $R$. It also cannot depend on the speed of light $c$. All dependence of $H^{(1,0+)}$
on $R$ and $c$ comes only through the relativistic effects, which are small corrections in the
low-$Z$ region. Therefore, numerical calculations of $H^{(1,0+)}$ performed with different choices
of $R$ and $c$ should have the same low-$Z$ limit.

\begin{figure}
\centerline{
\resizebox{0.5\textwidth}{!}{%
  \includegraphics{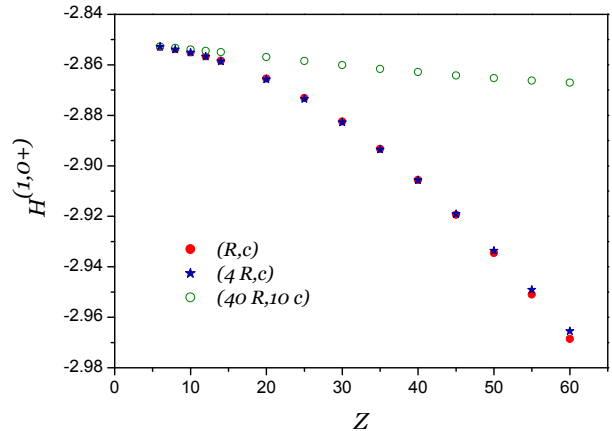}
}}
 \caption{(Color online)
The one-photon exchange fns correction to the bound-electron $g$ factor of the ground
state of Li-like ions, in terms of the function $H^{(1)}$ defined by Eq.~(\ref{eq:26}).
Numerical results for the actual values the nuclear charge radii and the speed of light $(R,c)$
(filled dots, red) are compared with the results obtained with with the nuclear charge radii
multiplied by a factor of 4 $(4R,c)$ (filled stars, blue) and with the results obtained with the
nuclear charge radii multiplied by a factor of 40 and the speed of light multiplied by 10
$(40R,10c)$ (open dots, green). \label{fig:H1}}
\end{figure}

The numerical results for the nuclear-size correction to the one-photon exchange are presented in
Table~\ref{tab:Z1} and shown graphically on Fig.~\ref{fig:H1}. We observe that the results obtained
with different values of $R$ and $c$ are in very good agreement for low $Z$. This agreement also
indicates that the results for $H^{(1)}$ are practically independent of the nuclear model.

%%%%%%%%%%%%%%%%%%%%%%%%%%%%%%%%%%%%%%%%%%%%%%%%%%%%%%%%%%%%%%%%%%%%%%%
\begin{table}
\caption{The one-photon exchange fns correction to the bound-electron $g$ factor of the ground state of Li-like
ions, in terms of the function $H^{(1)}$ defined by Eq.~(\ref{eq:26}). The column $(R,c)$ contains results obtained with
the actual values of the nuclear charge radii $R$ and the speed of light $c$. The column $(4R,c)$ presents results obtained with
the nuclear charge radii multiplied by a factor of 4. The column $(40R,10c)$ contains results obtained with
the nuclear charge radii multiplied by a factor of 40 and the speed of light multiplied by 10.
 \label{tab:Z1} }
\begin{center}
\begin{ruledtabular}
\begin{tabular}{l...}
%\hline\\[-0.5cm]
                $Z$
                & \multicolumn{1}{c}{$(R,c)$}
                & \multicolumn{1}{c}{$(4R,c)$}
                & \multicolumn{1}{c}{$(40R,10c)$}
 \\
\hline\\[-9pt]
%
%       &     (R,c)    &   (R*4,c)    &  (R*40,c*10)
%
   6   &    -2.x8529   &   -2.x8529    &   -2.x8527 \\
   8   &    -2.x8538   &   -2.x8539    &   -2.x8533 \\
   10  &    -2.x8550   &   -2.x8552    &   -2.x8539    \\
   12  &    -2.x8566   &   -2.x8568    &   -2.x8545       \\
   14  &    -2.x8584   &   -2.x8586    &   -2.x8550 \\
   20  &    -2.x8654   &   -2.x8658    &   -2.x8569    \\
   25  &    -2.x8731   &   -2.x8735    &   -2.x8585       \\
   30  &    -2.x8824   &   -2.x8828    &   -2.x8601          \\
   35  &    -2.x8933   &   -2.x8936    &   -2.x8616  \\
   40  &    -2.x9057   &   -2.x9057    &   -2.x8629 \\
   45  &    -2.x9194   &   -2.x9191    &   -2.x8642   \\
   50  &    -2.x9346   &   -2.x9336    &   -2.x8653      \\
   55  &    -2.x9510   &   -2.x9491    &   -2.x8663         \\
   60  &    -2.x9686   &   -2.x9655    &   -2.x8670            \\
\end{tabular}
\end{ruledtabular}
\end{center}
\end{table}

%%%%%%%%%%%%%%%%%%%%%%%%%%%%%%%%%%%%%%%%%%%%%%%%%%%%%%%%%%%%%%%%%%%%%%%%
%%%%%
%%%%%
%%%%%%%%%%%%%%%%%%%%%%%%%%%%%%%%%%%%%%%%%%%%%%%%%%%%%%%%%%%%%%%%%%%%%%%

The results obtained with enlarged speed of light show very weak $Z$ dependence, which might have
been anticipated since the $Z$ dependence of $H^{(1)}$ comes through the relativistic corrections
only. These results can be easily extrapolated to $Z\to 0$, yielding
\begin{align}
H^{(1,0)}= -2.8512\,(10)\,.
\end{align}
On the basis of this result, we conclude that the following weighted difference of the $2s$ and
$1s$ $g$-factors cancels most of the fns contribution of order $1/Z$ for light ions,
\begin{align}
\delta_{\Xi_1}g = \delta g^{(1)}(2s) - \Xi_0\, \left(-\frac{2851}{1000}\frac1Z\right) g^{(0)}(1s)\,.
\end{align}

\subsection{Two and more photon exchange fns correction}

The fns correction with two and more photon exchanges between the electrons is suppressed by the
factor of $1/Z^2$ with respect to the leading fns contributions. A parametrization of this term can
be given as
\begin{align} \label{eq:312}
 \delta g^{(2+)}_N = \delta g^{(0)}_N\, \frac1{Z^2} \,H^{(2+)}(\Za,R)\,,
\end{align}
where $\delta g^{(0)}_N$ is the one-electron nuclear-size correction defined in Eq.~(\ref{eq:01}), and
$H^{(2+)}$ is a slowly varying function of its arguments.

In order to compute the fns correction, we need to calculate the two and more photon exchange
correction for the extended and the point nucleus and take the difference,
\begin{align} \label{eq:313}
 \delta g^{(2+)}_N = \delta g^{(2+)}_{\rm ext} - \delta g^{(2+)}_{\rm pnt}\,.
\end{align}
In this work, we calculate $\delta g^{(2+)}_{\rm ext}$ and $\delta g^{(2+)}_{\rm pnt}$ within the
Breit approximation. The whole calculation is performed in three steps. In the first step, we solve
the no-pair Dirac-Coulomb-Breit Hamiltonian by the Configuration-Interaction Dirac-Fock-Sturm
(CI-DFS) method \cite{tupitsyn:03}. In the second step, we subtract the leading-order terms of
orders $1/Z^0$ and $1/Z^1$, thus identifying the contribution of order $1/Z^2$ and higher. The
subtraction terms of order $1/Z^0$ and $1/Z^1$ were calculated separately by perturbation theory.
In the third step, we repeat the calculation for the extended and the point nuclear models and, by
taking the difference, obtain the fns correction.

\begin{figure}
\centerline{
\resizebox{0.5\textwidth}{!}{%
  \includegraphics{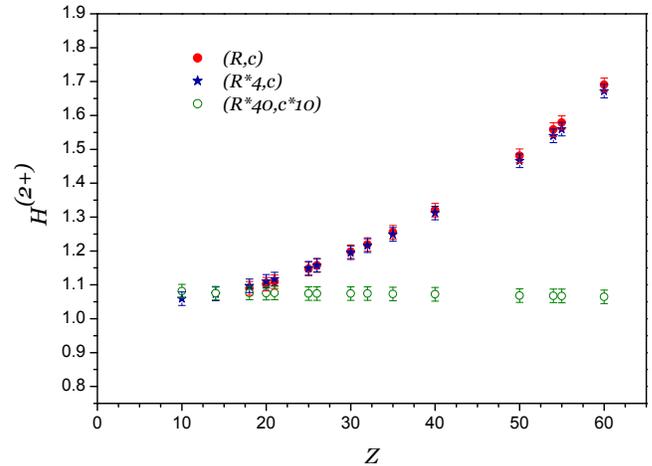}
}}
 \caption{(Color online)
The two and more photon exchange  fns correction to the bound-electron $g$ factor of the ground
state of Li-like ions, in terms of the function $H^{(2+)}$ defined by Eq.~(\ref{eq:312}).
Notations are the same as in Fig.~\ref{fig:H1}. \label{fig:H2}}
\end{figure}

The fns effect is very small in the low-$Z$ region, which makes it very difficult to obtain
reliable predictions for this correction. In order to be able to monitor the numerical accuracy, we
performed three sets of calculations. The first set $(R,c)$ was obtained with the actual values of
the nuclear charge radii $R$ and the speed of light $c$; the second set $(4R,c)$ was obtained with
the nuclear charge radii multiplied by a factor of 4; the third set $(40R,10c)$ was obtained with
the nuclear charge radii multiplied by a factor of 40 and the speed of light multiplied by 10. The
obtained results are listed in Table~\ref{tab:Z2} and presented in Fig.~\ref{fig:H2}.

Similarly to the one-photon exchange fns correction, we assume that the low-$Z$ limit of
$H^{(2+)}$, denoted as $H^{(2,0)}$, does not depend either on $R$ or on $c$. By extrapolating our
numerical results in Table~\ref{tab:Z2} to $Z\to 0$, we obtain the nonrelativistic value of the
$1/Z^2$ correction as
\begin{align} \label{eq:27}
H^{(2,0)}= 1.070\,(25)\,.
\end{align}
Based on this result, we conclude that for light ions, the following weighted difference of the $2s$
and $1s$ $g$-factors cancels most of the $1/Z^2$ fns contribution:
\begin{align} \label{eq:28}
\delta_{\Xi_2}g = \delta g^{(2+)}(2s) - \Xi_0\, \left(\frac{107}{100}\frac1{Z^2}\right) g^{(0)}(1s)\,.
\end{align}

%%%%%%%%%%%%%%%%%%%%%%%%%%%%%%%%%%%%%%%%%%%%%%%%%%%%%%%%%%%%%%%%%%%%%%%
\begin{table}
\caption{The two and more photon exchange  fns correction to the bound-electron $g$ factor of the ground state of Li-like
ions, in terms of the function $H^{(2+)}$ defined by Eq.~(\ref{eq:312}). Notations are the same as in Table~\ref{tab:Z1}.
 \label{tab:Z2} }
\begin{center}
\begin{ruledtabular}
\begin{tabular}{l...}
%\hline\\[-0.5cm]
                $Z$
                & \multicolumn{1}{c}{$(R,c)$}
                & \multicolumn{1}{c}{$(4R,c)$}
                & \multicolumn{1}{c}{$(40R,10c)$}
 \\
\hline\\[-9pt]
%
%       &     (R,c)    &   (R*4,c)    &  (R*40,c*10)
%
   10  &                & 1.x059\,(20)  & 1.x081\,(20) \\
   14  &                & 1.x073\,(20)  & 1.x075\,(20) \\
   20  & 1.x102\,(20)   & 1.x110\,(20)  & 1.x075\,(20)\\
   25  & 1.x157\,(20)   & 1.x149\,(20)  & 1.x074\,(20)\\
   30  & 1.x198\,(20)   & 1.x195\,(20)  & 1.x074\,(20)\\
   35  & 1.x255\,(20)   & 1.x249\,(20)  & 1.x073\,(20)\\
   40  & 1.x321\,(20)   & 1.x312\,(20)  & 1.x072\,(20)\\
%  45  & 1.x395   & 1.x409(???) & 1.x077 \\
   50  & 1.x481\,(20)   & 1.x466\,(20)  & 1.x068\,(20)\\
   55  & 1.x579\,(20)   & 1.x560\,(20)  & 1.x067\,(20)\\
   60  & 1.x690\,(20)   & 1.x672\,(20)  & 1.x064\,(20)\\
\end{tabular}
\end{ruledtabular}
\end{center}
\end{table}

%%%%%%%%%%%%%%%%%%%%%%%%%%%%%%%%%%%%%%%%%%%%%%%%%%%%%%%%%%%%%%%%%%%%%%%%
%%%%%
%%%%%
%%%%%%%%%%%%%%%%%%%%%%%%%%%%%%%%%%%%%%%%%%%%%%%%%%%%%%%%%%%%%%%%%%%%%%%

%%%%%%%%%%%%%%%%%%%%%%%%%%%%%%%%%%%%%%%%%%%%%%%%%%%%%%%%%%%%%%%%%%%%%%%%
%%%%%
%%%%%
%%%%%%%%%%%%%%%%%%%%%%%%%%%%%%%%%%%%%%%%%%%%%%%%%%%%%%%%%%%%%%%%%%%%%%%
\begin{figure*}[t]
\centerline{
\resizebox{\textwidth}{!}{%
  \includegraphics{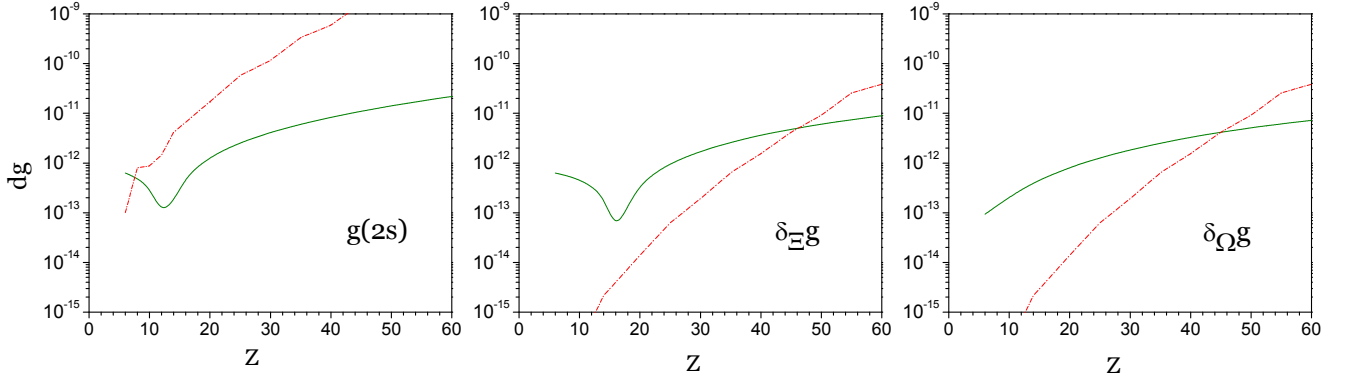}
}}
 \caption{(Color online) Comparison of the error $\delta g=({\partial g}/{\partial \alpha})\, \delta\alpha$
 due to the uncertainty of the fine-structure constant $\delta \alpha/\alpha = 3.2\times 10^{-10}$
 (solid line, green) and the error due to the  finite nuclear size effect (dashed-dot line, red), for the
 $g$-factor of the ground state of Li-like ions $g(2s)$ (left panel); for the weighted difference
 $\delta_{\Xi}g(Z)$ (middle panel); and for the weighted difference
 $\delta_{\Omega}g = \delta_{\Xi}g(Z) - \delta_{\Xi}g([Z/2])$ (right panel).
\label{fig:sens}}
\end{figure*}

%%%%%%%%%%%%%%%%%%%%%%%%%%%%%%%%%%%%%%%%%%%%%%%%%%%%%%%%%%%%%%%%%%%%%%%%
%%%%%
%%%%%
%%%%%%%%%%%%%%%%%%%%%%%%%%%%%%%%%%%%%%%%%%%%%%%%%%%%%%%%%%%%%%%%%%%%%%%
%\begin{figure}[t]
%\centerline{
%\resizebox{0.5\textwidth}{!}{%
%  \includegraphics{sensitivity1b.eps}
%}}
% \caption{(Color online) Comparison of the error $dg=\frac{\partial g}{\partial \alpha} \delta\alpha$ due to the uncertainty
% of the fine-structure constant $\delta \alpha/\alpha =
%3.2\times 10^{-10}$ (solid line, brown) and the error due to the
% finite nuclear size effect due to the model dependence and due to the uncertainty of the nuclear
% charge radii (dash-dotted, red).
%\label{fig:sens}}
%\end{figure}
%
%%%%%%%%%%%%%%%%%%%%%%%%%%%%%%%%%%%%%%%%%%%%%%%%%%%%%%%%%%%%%%%%%%%%%%%%
%%%%%
%%%%%
%%%%%%%%%%%%%%%%%%%%%%%%%%%%%%%%%%%%%%%%%%%%%%%%%%%%%%%%%%%%%%%%%%%%%%%
%\begin{figure}[t]
%\centerline{
%\resizebox{0.5\textwidth}{!}{%
%  \includegraphics{sensitivityM.eps}
%}}
% \caption{(Color online) The same as in Fig.~\ref{fig:sens}, but for the weighted difference
% $\delta_{\Omega}g = \delta_{\Xi}g(Z) - \delta_{\Xi}g(Z/2)$.
%\label{fig:sensM}}
%\end{figure}

\section{\label{sec:wdiff} The weighted difference of the $\bm {2s}$ and $\bm {1s}$ $\bm g$ factors}

Combining the results obtained in the previous section, we introduce the total $\Xi$-weighted
difference as follows
\begin{align} \label{eq:Xi}
\delta_{\Xi}g = g(2s) - \Xi\, g(1s)\,,
\end{align}
where $g(2s)$ is the $g$ factor of the ground state of the Li-like ion, $g(1s)$ is the  $g$ factor
of the ground state of the H-like ion, and the weight parameter $\Xi$ is defined by
Eq.~(\ref{eq:37}). Basing on the analysis of the preceding Section, we claim that in the
$\Xi$-weighted difference $\delta_{\Xi}g$, the nonrelativistic fns corrections to order $1/Z^0$,
$1/Z^1$, and $1/Z^2$ and, in addition, the relativistic contribution to order $(Z\alpha)^2/Z^0$ are
cancelled. A small remaining fns correction to $\delta_{\Xi}g$ is calculated numerically. The
definition of $\delta_{\Xi}g$ is based on the $\Za$ expansion of the fns corrections. Because of
this, it is applicable for low- and medium-$Z$ ions. For heavy systems, the $\Za$ expansion is no
longer useful. In this case, the cancellation of the fns effect in the weighted difference is still
possible but should be achieved differently \cite{shabaev:02:li,shabaev:06:prl}.

In Table~\ref{tab:1} we present the individual fns contributions to the $g$-factor of the ground state of
Li-like ions $g(2s)$, H-like ions $g(1s)$ and for the weighted difference $\delta_{\Xi}g$. We observe that
the uncertainty of the fns corrections for $g(2s)$ and $g(1s)$ is dominated by the nuclear-model and
nuclear-radii errors, which means they cannot be significantly improved. On the contrary, the fns effect
for $\delta_{\Xi}g$ is much smaller, and its uncertainty is mainly numerical, meaning that it can be
improved further.

%\newpage

%%%%%%%%%%%%%%%%%%%%%%%%%%%%%%%%%%%%%%%%%%%%%%%%%%%%%%%%%%%%%%%%%%%%%%%
\begin{longtable*}{ll...}
\caption{The fns corrections to the bound-electron $g$ factor of the ground state of Li-like and
H-like ions and their weighted difference, multiplied by a factor of $10^6$. The numbers in the
parentheses denote the uncertainty in the last figure. When three uncertainties are specified, the
first one is the numerical error, the second one the model-dependence error, and the third one the
uncertainty induced by the error of the nuclear charge radius. In the case only one uncertainty is
specified, it is the numerical error (whereas the other errors are significantly smaller and are
not indicated).
 \label{tab:1} }
\\ \colrule\hline
                $Z$ & Term
                & \multicolumn{1}{c}{$\delta g_{\rm N}(2s)$}
                & \multicolumn{1}{c}{$\Xi_i/Z^i\, \delta g_{\rm N}(1s)$}
                & \multicolumn{1}{c}{$\delta g_{\rm N}(2s) - \Xi_i/Z^i\, \delta g_{\rm N}(1s)$}
 \\ \hline\\[-5pt]
  6 &  $1/Z^0$    & 0.000\,x050\,99\,(0)(1)(9) & 0.000\,x050\,99\,(0)(1)(9) & 0.\,x \\
    &  $\alpha/Z^0$    & -0.000\,x000\,05\,(2) & -0.000\,x000\,071\,(2) & 0.000\,x000\,03\,(2) \\
    &  $1/Z^1$    & -0.000\,x024\,24\,(0)(0)(4) & -0.000\,x024\,23\,(0)(0)(4) & -0.000\,x000\,016\,(1)(0)(0) \\
    &  $1/Z^{2+}$ & 0.000\,x001\,52\,(4) & 0.000\,x001\,515 &  0.000\,x000\,00\,(4) \\
    &  Total      & 0.000\,x028\,2\,(0)(0)(1) & 0.000\,x0282\,(0)(0)(1) & 0.000\,x000\,01\,(4)(0)(0) \\  \\
  8 &  $1/Z^0$    & 0.000\,x194\,7\,(0)(0)(7) & 0.000\,x194\,7\,(0)(0)(7) & 0.\,x \\
    &  $\alpha/Z^0$    & -0.000\,x000\,25\,(3) & -0.000\,x000\,317\,(5) & 0.000\,x000\,07\,(3) \\
    &  $1/Z^1$    & -0.000\,x069\,5\,(0)(0)(3) & -0.000\,x069\,4\,(0)(0)(3) & -0.000\,x000\,068\,(1)(0)(0) \\
    &  $1/Z^{2+}$ & 0.000\,x003\,26\,(8) & 0.000\,x003\,256 &  0.000\,x000\,00\,(8) \\
    &  Total      & 0.000\,x128\,3\,(1)(0)(8) & 0.000\,x128\,3\,(0)(0)(8) & 0.000\,x000\,00\,(8)(0)(0) \\  \\
 10 &  $1/Z^0$    & 0.000\,x598\,3\,(0)(1)(8) & 0.000\,x598\,3\,(0)(1)(8) & -0.000\,x000\,002 \\
    &  $\alpha/Z^0$    & -0.000\,x000\,90\,(8) & -0.000\,x001\,12\,(1) & 0.000\,x000\,22\,(8) \\
    &  $1/Z^1$    & -0.000\,x170\,8\,(0)(0)(2) & -0.000\,x170\,6\,(0)(0)(2) & -0.000\,x000\,241\,(1)(0)(0) \\
    &  $1/Z^{2+}$ & 0.000\,x006\,4\,(1) & 0.000\,x006\,40 &  0.000\,x000\,0\,(1) \\
    &  Total      & 0.000\,x433\,0\,(2)(1)(9) & 0.000\,x433\,0\,(0)(1)(9) &  0.000\,x000\,0\,(2)(0)(0) \\  \\
 12 &  $1/Z^0$    & 0.001\,x307\,(0)(0)(1) & 0.001\,x307\,(0)(0)(1) & -0.000\,x000\,007 \\
    &  $\alpha/Z^0$    & -0.000\,x002\,3\,(2) & -0.000\,x002\,74\,(2) & 0.000\,x000\,4\,(2) \\
    &  $1/Z^1$    & -0.000\,x311\,1\,(0)(1)(3) & -0.000\,x310\,5\,(0)(1)(3) & -0.000\,x000\,604\,(1)(0)(1) \\
    &  $1/Z^{2+}$ & 0.000\,x009\,7\,(2) & 0.000\,x009\,71 &  0.000\,x000\,0\,(2) \\
    &  Total      & 0.001\,x003\,(0)(0)(1) & 0.001\,x003\,(0)(0)(1) & -0.000\,x000\,2\,(3)(0)(0) \\  \\
 14 &  $1/Z^0$    & 0.002\,x580\,(0)(1)(4) & 0.002\,x580\,(0)(1)(4) & -0.000\,x000\,026\,(0)(1)(0) \\
    &  $\alpha/Z^0$    & -0.000\,x005\,1\,(3) & -0.000\,x005\,96\,(3) & 0.000\,x000\,8\,(3) \\
    &  $1/Z^1$    & -0.000\,x5267\,(0)(2)(8) & -0.000\,x525\,3\,(0)(2)(8) & -0.000\,x001\,353\,(1)(0)(2) \\
    &  $1/Z^{2+}$ & 0.000\,x014\,1\,(3) & 0.000\,x014\,1 &  0.000\,x000\,0\,(3) \\
    &  Total      & 0.002\,x062\,(0)(1)(4) & 0.002\,x062\,(0)(1)(4) & -0.000\,x000\,6\,(4)(0)(0) \\  \\
 20 &  $1/Z^0$    & 0.014\,x41\,(0)(1)(2) & 0.014\,x41\,(0)(1)(2) & -0.000\,x000\,554\,(0)(7)(1) \\
    &  $\alpha/Z^0$    & -0.000\,x038\,(2) & -0.000\,x041\,4\,(1) & 0.000\,x004\,(2) \\
    &  $1/Z^1$    & -0.002\,x064\,(0)(1)(2) & -0.002\,x054\,(0)(1)(2) & -0.000\,x010\,31\,(0)(0)(1) \\
    &  $1/Z^{2+}$ & 0.000\,x040\,0\,(7) & 0.000\,x038\,5 & 0.000\,x001\,4\,(7) \\
    &  Total      & 0.012\,x34\,(0)(1)(2) & 0.012\,x35\,(0)(1)(2) & -0.000\,x006\,(2)(0)(0) \\  \\
 25 &  $1/Z^0$    & 0.043\,x36\,(0)(3)(5) & 0.043\,x36\,(0)(3)(5) & -0.000\,x003\,90\,(0)(4)(1) \\
    &  $\alpha/Z^0$    & -0.000\,x136\,(5) & -0.000\,x141\,4\,(2) & 0.000\,x005\,(5) \\
    &  $1/Z^1$    & -0.004\,x983\,(0)(3)(6) & -0.004\,x945\,(0)(3)(6) & -0.000\,x037\,92\,(0)(2)(4) \\
    &  $1/Z^{2+}$ & 0.000\,x080\,(1) & 0.000\,x074 & 0.000\,x006\,(1) \\
    &  Total      & 0.038\,x32\,(1)(3)(5) & 0.038\,x35\,(0)(3)(5) & -0.000\,x031\,(5)(0)(0) \\  \\
 30 &  $1/Z^0$    & 0.111\,x34\,(0)(8)(8) & 0.111\,x36\,(0)(8)(8) & -0.000\,x020\,3\,(0)(1)(0) \\
    &  $\alpha/Z^0$    & -0.000\,x39\,(1) & -0.000\,x398\,9\,(5) & 0.000\,x01\,(1) \\
    &  $1/Z^1$    & -0.010\,x697\,(0)(8)(8) & -0.010\,x583\,(0)(8)(8) & -0.000\,x114\,72\,(0)(9)(9) \\
    &  $1/Z^{2+}$ & 0.000\,x148\,(2) & 0.000\,x132 & 0.000\,x016\,(2) \\
    &  Total      & 0.100\,x40\,(1)(8)(8) & 0.100\,x51\,(0)(8)(8) & -0.000\,x11\,(1)(0)(0) \\  \\
 35 &  $1/Z^0$    & 0.258\,x8\,(0)(2)(3) & 0.258\,x9\,(0)(2)(3) & -0.000\,x086\,4\,(0)(5)(1) \\
    &  $\alpha/Z^0$    & -0.000\,x97\,(2) & -0.000\,x995\,4\,(6) & 0.000\,x02\,(2) \\
    &  $1/Z^1$    & -0.021\,x40\,(0)(2)(2) & -0.021\,x09\,(0)(2)(2) & -0.000\,x305\,7\,(0)(3)(3) \\
    &  $1/Z^{2+}$ & 0.000\,x265\,(4) & 0.000\,x226 & 0.000\,x039\,(4) \\
    &  Total      & 0.236\,x7\,(0)(2)(3) & 0.237\,x1\,(0)(2)(3) & -0.000\,x33\,(2)(0)(0) \\  \\
 40 &  $1/Z^0$    & 0.527\,x6\,(0)(5)(2) & 0.527\,x9\,(0)(5)(2) & -0.000\,x298\,(0)(1)(0) \\
    &  $\alpha/Z^0$    & -0.002\,x10\,(3) & -0.002\,x125\,(1) & 0.000\,x03\,(3) \\
    &  $1/Z^1$    & -0.038\,x33\,(0)(4)(2) & -0.037\,x63\,(0)(4)(2) & -0.000\,x699\,6\,(0)(8)(3) \\
    &  $1/Z^{2+}$ & 0.000\,x436\,(7) & 0.000\,x353 & 0.000\,x083\,(7) \\
    &  Total      & 0.487\,x6\,(0)(5)(2) & 0.488\,x5\,(0)(5)(2) & -0.000\,x89\,(3)(0)(0) \\  \\
 45 &  $1/Z^0$    & 1.076\,x\,(0)(1)(1) & 1.077\,x\,(0)(1)(1) & -0.000\,x982\,(0)(3)(1) \\
    &  $\alpha/Z^0$    & -0.004\,x47\,(3) & -0.004\,x486\,(3) & 0.000\,x02\,(4) \\
    &  $1/Z^1$    & -0.069\,x81\,(0)(8)(7) & -0.068\,x24\,(0)(8)(7) & -0.001\,x574\,(0)(2)(2) \\
    &  $1/Z^{2+}$ & 0.000\,x74\,(1) & 0.000\,x569 & 0.000\,x17\,(1) \\
    &  Total      & 1.003\,x\,(0)(1)(1) & 1.005\,x\,(0)(1)(1) & -0.002\,x37\,(4)(0)(0) \\  \\
 50 &  $1/Z^0$    & 2.050\,x\,(0)(3)(2) & 2.053\,x\,(0)(3)(2) & -0.002\,x885\,(0)(7)(3) \\
    &  $\alpha/Z^0$    & -0.008\,x73\,(5) & -0.008\,x684\,(5) & -0.000\,x05\,(5) \\
    &  $1/Z^1$    & -0.120\,x3\,(0)(2)(1) & -0.117\,x1\,(0)(1)(1) & -0.003\,x262\,(0)(5)(3) \\
    &  $1/Z^{2+}$ & 0.001\,x21\,(2) & 0.000\,x878 & 0.000\,x34\,(2) \\
    &  Total      & 1.922\,x\,(0)(3)(2) & 1.928\,x\,(0)(3)(2) & -0.005\,x86\,(5)(1)(0) \\  \\
 55 &  $1/Z^0$    & 3.788\,x\,(0)(5)(7) & 3.796\,x\,(0)(5)(7) & -0.007\,x95\,(0)(1)(2) \\
    &  $\alpha/Z^0$    & -0.016\,x29\,(9) & -0.016\,x037\,(9) & -0.000\,x26\,(9) \\
    &  $1/Z^1$    & -0.203\,x2\,(0)(3)(4) & -0.196\,x8\,(0)(3)(3) & -0.006\,x47\,(0)(1)(1) \\
    &  $1/Z^{2+}$ & 0.001\,x98\,(3) & 0.001\,x342 & 0.000\,x63\,(3) \\
    &  Total      & 3.570\,x\,(0)(5)(7) & 3.584\,x\,(0)(5)(7) & -0.014\,x05\,(9)(2)(2) \\  \\
 60 &  $1/Z^0$    & 6.74\,x(0)(1)(1) & 6.76\,x(0)(1)(1) & -0.020\,x51\,(0)(2)(2) \\
    &  $\alpha/Z^0$    & -0.028\,x7\,(2) & -0.027\,x96\,(2) & -0.000\,x8\,(2) \\
    &  $1/Z^1$    & -0.333\,x6\,(0)(5)(3) & -0.321\,x4\,(0)(5)(3) & -0.012\,x24\,(0)(2)(1) \\
    &  $1/Z^{2+}$ & 0.003\,x17\,(4) & 0.002\,x010 & 0.001\,x16\,(4) \\
    &  Total      & 6.38\,x(0)(1)(1) & 6.42\,x(0)(1)(1) & -0.032\,x4\,(2)(0)(0)
 \\
\\ \colrule\hline
\end{longtable*}

We would like now to address the question whether the weighted difference $\delta_{\Xi}g$ might be
useful for the determination of the fine-structure constant $\alpha$. The leading dependence of
$\delta_{\Xi}g$ on $\alpha$ is given by the expansion
\begin{align}\label{eq42}
\delta_{\Xi}g =
2\,(1-\Xi) -\frac23(\Za)^2\left(\frac14-\Xi\right) + \frac{\alpha}{\pi}(1-\Xi) + \ldots\,,
\end{align}
where the second term in the right-hand-side stems from the binding corrections, whereas the third
term is due to the one-loop free-electron QED effect. In the above equation, we keep $\Xi$ fixed,
ignoring its dependence on $\alpha$, since it does not contribute to the sensitivity of
$\delta_{\Xi}g$ on $\alpha$ (the same value of $\Xi$ should be used when comparing the experimental
and theoretical values of $\delta_{\Xi}g$). By varying $\alpha$ in Eq.~(\ref{eq42}) within its
current error bars of $\delta \alpha/\alpha = 3.2\times 10^{-10}$ \cite{mohr:12:codata}, the
corresponding error of $\delta_{\Xi}g$ can be obtained.

In Fig.~\ref{fig:sens} we compare the uncertainty due to $\alpha$ and the uncertainty due to the
nuclear model and radius, keeping in mind that the latter defines the ultimate limit of the
accuracy of theoretical calculations. The left panel of Fig.~\ref{fig:sens} shows this comparison
for the $g$-factor of the ground state of Li-like ions $g(2s)$, whereas the middle panel gives the
same comparison for the $\Xi$-weighted difference $\delta_{\Xi}g$. The dip of the
$\alpha$-sensitivity curve around $Z = 16$ is caused by the fact that the dependence of the binding
and the free-QED effects on $\alpha$ in Eq.~(\ref{eq42}) (second and third terms) have different
signs, and thus cancel each other in this $Z$ region. From Fig.~\ref{fig:sens} we can conclude that
up to $Z \approx 45$, the weighted difference $\delta_{\Xi}g$ yields possibilities for an improved
determination of $\alpha$.

The determination of $\alpha$ from $\delta_{\Xi}g$ has two drawbacks. The first one is the
cancellation of $\alpha$ dependence of $\delta_{\Xi}g$ around $Z = 16$, leading to a loss of
sensitivity to $\alpha$ in this $Z$ region. The second one is that $\delta_{\Xi}g$ contains the
same free-QED part which is used for the determination of $\alpha$ from the free-electron $g$
factor, which means that these two determinations cannot be regarded as fully independent. Both
drawbacks can be avoided by introducing another difference,
\begin{align} \label{eq:Omega}
\delta_{\Omega}g = \delta_{\Xi}g(Z) - \delta_{\Xi}g([Z/2])\,,
\end{align}
with $\delta_{\Xi}g(Z)$ being the weighted difference (\ref{eq:Xi}) for the nuclear charge $Z$, and
$\delta_{\Xi}g([Z/2])$ is the corresponding difference for the nuclear charge $[Z/2]$, where
$[\ldots]$ stands for the upper or the lower integer part. In the difference $\delta_{\Omega}g$,
most free-QED contributions vanish. So, by a small sacrifice of the sensitivity of the binding
effects to $\alpha$, we removed the dip around $Z = 16$ and made the theory of the weighted
difference (almost) independent on the theory of the free-electron $g$-factor.

The right panel of Fig~\ref{fig:sens} presents the comparison of the uncertainty due to $\alpha$ with the
error of the fns effect for the weighted difference $\delta_{\Omega}g$. One finds a smooth dependence of the
sensitivity to $\alpha$ on $Z$, without any dip in the region around $Z = 16$. We observe that in the region
$Z = 10 - 20$, the weighted difference $\delta_{\Omega}g$ offers better possibilities for determining
$\alpha$ than $\delta_{\Xi}g$.

Employing the difference $\delta_{\Omega}g$ can be also advantageous from the experimental point of view. It
can be rewritten as
\begin{eqnarray} \label{eq:Omega-rewritten}
\delta_{\Omega}g &=& g(2s,Z)-g(2s,Z_2) \\
                 &-& \Xi(Z) \left[g(1s,Z)-g(1s,Z_2) \right] \nonumber \\
                 &-& g(1s,Z_2) \left[\Xi(Z)-\Xi(Z_2) \right] \nonumber \,,
\end{eqnarray}
with $Z_2 = [Z/2]$. We thus observe that $\delta_{\Omega}g$ can be effectively determined in an
experiment by measuring two equal-weight $g$-factor differences (namely, the ones in the first and
second rows of the above equation) and $g(1s,Z_2)$. The equal-weight differences may be measured
with largely suppressed systematic errors and thus can be determined in near-future experiments
much more accurately than the $g$-factors of individual ions. The last term in
Eq.~(\ref{eq:Omega-rewritten}) is suppressed by a small factor of $[\Xi(Z)-\Xi(Z_2)] \approx
0.02-0.04$ in the region of interest. Therefore, the experimental error of $\delta_{\Omega}g$ can
be significantly improved as compared to that of the absolute $g$-factors.

Let us now turn to the experimental consequences of the present calculations. So far, the only
element for which the weighted difference $\delta_{\Xi}g$ has been measured is silicon. In
Table~\ref{tab:si} we collect the individual theoretical contributions to
$\delta_{\Xi}g(^{29}\mbox{\rm Si})$. Theoretical results for various effects were taken from the
literature,
Refs.~\cite{yerokhin:04,glazov:04:pra,pachucki:05:gfact,volotka:09,glazov:10,volotka:14,aoyama:15}.
The total theoretical value is compared to the experimental result
\cite{sturm:13:Si,wagner:13,sturm:11}. The errors of the Dirac value and of the one-loop free QED
($\sim \alpha (\Za)^0$) result specified in the table are due to the uncertainty of the current
value of $\alpha^{-1} = 137.035\,999\,074\,(44)\,$ \cite{mohr:12:codata}. The uncertainty of the
fns effect specified in the table is $6\times 10^{-13}$, which is already smaller than the
uncertainty of the Dirac value due to $\alpha$. The fns uncertainty is of purely numerical origin,
i.e. it does not influenced by the errors due to the rms charge radius and the nuclear charge
distribution, and thus it can be further improved in future calculations.

Table~\ref{tab:si} illustrates another advantage of the $\Xi$-weighted difference: the
contributions of one-electron binding QED effects to $\delta_{\Xi}g$ are much smaller than those to
$g(2s)$. This is explained by the fact that these effects largely originate from short distances,
similarly to the fns effect, and thus are significantly canceled in the difference. In particular,
the uncertainty of $\delta_{\Xi}g(\mbox{\rm Si})$ due to three-loop binding QED effects is on the
$10^{-12}$ level, implying that these effects do not need to be known to a high degree of accuracy
for the determination of $\alpha$.

Table~\ref{tab:si} shows that the present experimental and theoretical precision of
$\delta_{\Xi}g({\rm Si})$ is on the level of few parts in $10^{-9}$, which is significantly worse
than the precision achieved for other systems (in particular, H-like carbon, where the present
experimental and theoretical uncertainties are, correspondingly, $6\times 10^{-11}$ and $6\times
10^{-12}$ \cite{sturm:14}). This underperformance is, however, more due to a lack of motivation
than due to principal obstacles.

On the experimental side, the same precision as for H-like carbon can be also obtained for
$\delta_{\Xi}g({\rm C})$, with an existing ion trap \cite{sturm:priv}. Further experimental advance
is anticipated that could bring one or two orders of magnitude of improvement \cite{sturm:priv}. On
the theoretical side, the modern nonrelativistic quantum electrodynamics (NRQED) approach (see,
e.g., \cite{puchalski:14}) can apparently provide a theoretical result for Li-like carbon with the
same accuracy as obtained for its H-like counterpart \cite{pachucki:priv}. Moreover, further
theoretical advance is possible: the two-loop QED corrections of order $\alpha^2(\Za)^5$ and the
three-loop QED corrections of order $\alpha^3(\Za)^4$ can be calculated, both for H-like and
Li-like ions \cite{pachucki:priv}.

As we are presently interested in light ions, the best way for the advancement of theory would be a
combination of two complementary methods. The first one is the NRQED method (used, e.g., in
\cite{yan:01:prl}) that accounts for the nonrelativistic electron-electron interactions to all
orders in $1/Z$, but expands the QED and relativistic effects in powers of $\alpha$ and $\Za$. The
second approach (used, e.g., in \cite{glazov:04:pra,volotka:09,glazov:10,volotka:14}) accounts for
the relativistic effects to all orders in $\Za$ but employs perturbation expansions in $\alpha$
(QED effects) and in $1/Z$ (electron-electron interaction). Matching the coefficients of the $\Za$
and $1/Z$ expansions from the two methods allows one to combine them together, as it was done for
energy levels in Ref.~\cite{yerokhin:10:helike}. As a result of this procedure, only higher-order
corrections in $\Za$ will be expanded in $1/Z$ and only higher-order corrections in $1/Z$ will be
expanded in $\Za$. This approach should allow one to advance theory to the level required for a
determination of $\alpha$.

The principal limitation for the theory is set by the non-trivial nuclear structural effects, such
as the nuclear deformation, nuclear polarization, etc. For light ions, the leading nuclear effects
are described by effective operators proportional to the Dirac delta function $\delta(\bm{r})$.
Such effects are canceled in the weighted difference $\delta_{\Xi}g$. We estimate that the
uncertainty due to the remaining nuclear effects in $\delta_{\Xi}g$ should be of the same order as
the nuclear-model dependence error of the fns effect. From the breakdown in Table~\ref{tab:1} we
deduce that for silicon, this error is by about two orders of magnitude smaller than the
uncertainty due to $\alpha$. We thus conclude that the nuclear effects do not represent any
obstacles for the determination of $\alpha$ from $\delta_{\Xi}g$ and $\delta_{\Omega}g$.

%%%%%%%%%%%%%%%%%%%%%%%%%%%%%%%%%%%%%%%%%%%%%%%%%%%%%%%%%%%%%%%%%%%%%%%
\begin{table*}
\caption{Individual contributions to the weighted difference $\delta_{\Xi}g$ for $^{29}$Si,
$M/m = 52806.93396$, $\Xi = 0.101136233077060$. \label{tab:si}}
\begin{center}
%\begin{ruledtabular}
\begin{tabular}{ll.}
\hline
\hline\\[-0.25cm]
                \multicolumn{1}{l}{Contribution}
                & \multicolumn{1}{l}{Order}
                & \multicolumn{1}{c}{Value}
 \\
\hline\\[-9pt]
    Dirac        &&      1.796\,687\,x854\,216\,5\,(7) \\[2pt]
  $1$-loop QED
 &   $\alpha(\Za)^0$     &     0.002\,087\,x898\,255\,0\,(7)   \\
&    $\alpha(\Za)^2$     &     0.000\,000\,x601\,506\,0     \\
&    $\alpha(\Za)^4$     &     0.000\,000\,x014\,797\,0     \\
&    $\alpha(\Za)^{5+}$  &     0.000\,000\,x015\,48\,(52)   \\[2pt]
  $2$-loop QED
  &  $\alpha^2(\Za)^0$     &    -0.000\,003\,x186\,116\,6    \\
  &  $\alpha^2(\Za)^2$     &    -0.000\,000\,x000\,917\,9    \\
  &  $\alpha^2(\Za)^4$     &    -0.000\,000\,x000\,084\,4    \\
  &  $\alpha^2(\Za)^{5+}$  &     0.000\,000\,x000\,00\,(13)    \\[2pt]
  $\geq 3$-loop QED
  & $\alpha^{3+}(\Za)^0$     &     0.000\,000\,x026\,514\,9\,(1)   \\
  &  $\alpha^{3+}(\Za)^2$    &     0.000\,000\,x000\,007\,6      \\
  &  $\alpha^{3+}(\Za)^{4+}$ &     0.000\,000\,x000\,000\,0\,(11)   \\[2pt]
   Recoil                 &
   $m/M(\Za)^{2+}$         &     0.000\,000\,x029\,4\,(10)     \\[2pt]
% ---
% Two-electron corrections:
    1-photon exchange&
 $(1/Z)(\Za)^{2+}$          &  0.000\,321\,x590\,803\,3   \\
    2-photon exchange&
 $(1/Z^2)(\Za)^{2+}$        & -0.000\,006\,x876\,0\,(5)     \\
$\geq 3$-photon exchange &
 $(1/Z^{3+})(\Za)^{2+}$        &  0.000\,000\,x093\,0\,(60)    \\[2pt]
    2-electron QED   &
 $(\alpha/Z)(\Za)^{2+}$   & -0.000\,000\,x236\,0\,(50)     \\
    2-electron Recoil&
   $(m/M)(1/Z)(\Za)^{2+}$& -0.000\,000\,x011\,6\,(7)      \\[2pt]
% ---
Finite nuclear size &&  -0.000\,000\,x000\,000\,6\,(4)  \\[2pt]
 %--------------------
\hline\\[-9pt]
    Total theory    &&   1.799\,087\,x813\,2\,(79)     \\
    Experiment  \cite{sturm:13:Si,wagner:13}
                    &&   1.799\,087\,x812\,5\,(21)   \\
\hline
\hline\\[-0.5cm]
\end{tabular}
%\end{ruledtabular}
\end{center}
\end{table*}

\section{Conclusion}

In this work we investigated specific weighted differences of the $g$-factors of H- and Li-like
ions of the same element. An accurate formula was obtained for the weight parameter $\Xi$,
determined by requiring cancellation of the nonrelativistic finite nuclear size corrections to
orders $1/Z^0$, $1/Z^1$, and $1/Z^2$ and, in addition, the relativistic contribution to order
$(Z\alpha)^2/Z^0$. The coefficients of the $\Za$ expansion of the finite nuclear size corrections
were obtained by performing accurate numerical calculations and fitting the results to the known
expansion form. It was demonstrated that the $\Xi$- and $\Omega$-weighted differences, as given by
Eqs.~(\ref{eq:Xi}) and (\ref{eq:Omega}), can be used for an efficient suppression of nuclear
effects. The residual uncertainty due to nuclear effects is smaller than the uncertainty due to the
currently accepted value of the fine-structure constant $\alpha$. The $\Xi$- and $\Omega$-weighted
differences may be used in future to determine $\alpha$ from a comparison of theoretical and
experimental bound-electron $g$-factors with an accuracy competitive with or better than the
present literature value.

\begin{acknowledgements}

V.A.Y. and Z.H. acknowledge helpful conversations with Sven Sturm. V.A.Y. acknowledges support by
the Ministry of Education and Science of the Russian Federation (program for organizing and
carrying out scientific investigations) and by RFBR (grant No. 16-02-00538). E.B. acknowledges
support from G-RISC, project No. P-2014a-9.

\end{acknowledgements}

%\bibliographystyle{../bibtex/phaip30}
%\bibliography{../bibtex/hfst}

\begin{thebibliography}{10}

\bibitem{sturm:14} S.~Sturm, F.~K\"ohler, J.~Zatorski, A.~Wagner, Z.~Harman, G.~Werth, W.~Quint,
  C.~H. Keitel, and K.~Blaum,
\newblock Nature {\bf 506}, 467–470 (2014).

\bibitem{wagner:13} A.~Wagner, S.~Sturm, F.~K\"ohler, D.~A. Glazov, A.~V. Volotka, G.~Plunien,
  W.~Quint, G.~Werth, V.~M. Shabaev, and K.~Blaum,
\newblock Phys. Rev. Lett. {\bf 110}, 033003 (2013).

\bibitem{sturm:priv} S.~Sturm,
\newblock Private communication, 2015.

\bibitem{mohr:12:codata} P.~J. Mohr, B.~N. Taylor, and D.~B. Newell,
\newblock Rev. Mod. Phys. {\bf 84}, 1527 (2012).

\bibitem{shabaev:02:li} V.~M. Shabaev, D.~A. Glazov, M.~B. Shabaeva, V.~A. Yerokhin, G.~Plunien,
    and
  G.~Soff,
\newblock Phys. Rev. A {\bf 65}, 062104 (2002).

\bibitem{shabaev:06:prl} V.~M. Shabaev, D.~A. Glazov, N.~S. Oreshkina, A.~V. Volotka, G.~Plunien,
    H.-J.
  Kluge, and W.~Quint,
\newblock Phys. Rev. Lett. {\bf 96}, 253002 (2006).

\bibitem{bouchendira:11} R.~Bouchendira, P.~Clad\'e, S.~Guellati-Kh\'elifa, F.~m.~c. Nez, and
    F.~m.~c.
  Biraben,
\newblock Phys. Rev. Lett. {\bf 106}, 080801 (2011).

\bibitem{aoyama:12} T.~Aoyama, M.~Hayakawa, T.~Kinoshita, and M.~Nio,
\newblock Phys. Rev. Lett. {\bf 109}, 111807 (2012).

\bibitem{aoyama:15} T.~Aoyama, M.~Hayakawa, T.~Kinoshita, and M.~Nio,
\newblock Phys. Rev. D {\bf 91}, 033006 (2015).

\bibitem{yerokhin:16:gfact:prl} V.~A. Yerokhin, E.~Berseneva, Z.~Harman, I.~I. Tupitsyn, and C.~H.
    Keitel,
\newblock Phys. Rev. Lett. {\bf 116}, 100801 (2016).

\bibitem{rose:61} M.~E. Rose,
\newblock {\em {Relativistic Electron Theory}},
\newblock John Wiley \& Sons, NY, 1961.

\bibitem{karshenboim:05} S.~G. Karshenboim, R.~N. Lee, and A.~I. Milstein,
\newblock Phys. Rev. A {\bf 72}, 042101 (2005).

\bibitem{shabaev:93:fns} V.~M. Shabaev,
\newblock J. Phys. B {\bf 26}, 1103 (1993).

\bibitem{karshenboim:00:pla} S.~G. Karshenboim,
\newblock Phys. Lett. {\bf A 266}, 380  (2000).

\bibitem{glazov:01:pla} D.~A. Glazov and V.~M. Shabaev,
\newblock Phys. Lett. A {\bf 297}, 408 (2002).

\bibitem{shabaev:04:DKB} V.~M. Shabaev, I.~I. Tupitsyn, V.~A. Yerokhin, G.~Plunien, and G.~Soff,
\newblock Phys. Rev. Lett. {\bf 93}, 130405 (2004).

\bibitem{angeli:13} I.~Angeli and K.~Marinova,
\newblock At. Data Nucl. Data Tabl. {\bf 99}, 69  (2013).

\bibitem{yerokhin:13:jpb} V.~A. Yerokhin, C.~H. Keitel, and Z.~Harman,
\newblock J. Phys. B {\bf 46}, 245002 (2013).

\bibitem{yerokhin:10:sehfs} V.~A. Yerokhin and U.~D. Jentschura,
\newblock Phys. Rev. A {\bf 81}, 012502 (2010).

\bibitem{shabaev:03:PSAS} V.~Shabaev,
\newblock in {\em Precision Physics of Simple Atomic Systems}, ed. S.~G.
  Karshenboim and V.~B. Smirnov, (Lecture Notes in Physics, Berlin, 2003, Springer), p. 97.

\bibitem{johnson:88} W.~R. Johnson, S.~A. Blundell, and J.~Sapirstein,
\newblock Phys. Rev. A {\bf 37}, 307  (1988).

\bibitem{tupitsyn:03} I.~I. Tupitsyn, V.~M. Shabaev, J.~R. Crespo L\'opez-Urrutia,
  I.~Dragani\ifmmode~\acute{c}\else \'{c}\fi{}, R.~Soria~Orts, and J.~Ullrich,
\newblock Phys. Rev. A {\bf 68}, 022511 (2003).

\bibitem{yerokhin:04} V.~A. Yerokhin, P.~Indelicato, and V.~M. Shabaev,
\newblock Phys. Rev. A {\bf 69}, 052503 (2004).

\bibitem{glazov:04:pra} D.~A. Glazov, V.~M. Shabaev, I.~I. Tupitsyn, A.~V. Volotka, V.~A. Yerokhin,
  G.~Plunien, and G.~Soff,
\newblock Phys. Rev. A {\bf 70}, 062104 (2004).

\bibitem{pachucki:05:gfact} K.~Pachucki, A.~Czarnecki, U.~D. Jentschura, and V.~A. Yerokhin,
\newblock Phys. Rev. A {\bf 72}, 022108 (2005).

\bibitem{volotka:09} A.~V. Volotka, D.~A. Glazov, V.~M. Shabaev, I.~I. Tupitsyn, and G.~Plunien,
\newblock Phys. Rev. Lett. {\bf 103}, 033005 (2009).

\bibitem{glazov:10} D.~A. Glazov, A.~V. Volotka, V.~M. Shabaev, I.~I. Tupitsyn, and G.~Plunien,
\newblock Phys. Rev. A {\bf 81}, 062112 (2010).

\bibitem{volotka:14} A.~V. Volotka, D.~A. Glazov, V.~M. Shabaev, I.~I. Tupitsyn, and G.~Plunien,
\newblock Phys. Rev. Lett. {\bf 112}, 253004 (2014).

\bibitem{sturm:13:Si} S.~Sturm, A.~Wagner, M.~Kretzschmar, W.~Quint, G.~Werth, and K.~Blaum,
\newblock Phys. Rev. A {\bf 87}, 030501 (2013).

\bibitem{sturm:11} S.~Sturm, A.~Wagner, B.~Schabinger, J.~Zatorski, Z.~Harman, W.~Quint, G.~Werth,
  C.~H. Keitel, and K.~Blaum,
\newblock Phys. Rev. Lett. {\bf 107}, 023002 (2011).

\bibitem{puchalski:14} M.~Puchalski and K.~Pachucki,
\newblock Phys. Rev. Lett. {\bf 113}, 073004 (2014).

\bibitem{pachucki:priv} K.~Pachucki,
\newblock Private communication, 2015.

\bibitem{yan:01:prl} Z.-C. Yan,
\newblock Phys. Rev. Lett. {\bf 86}, 5683  (2001).

\bibitem{yerokhin:10:helike} V.~A. Yerokhin and K.~Pachucki,
\newblock Phys. Rev. A {\bf 81}, 022507 (2010).

\end{thebibliography}

\end{document}